\documentclass[pra,aps,amsmath,reprint,10pt,longbibliography]{revtex4-1}

\usepackage[utf8]{inputenc}
\usepackage{dcolumn}
\usepackage{amsbsy}
\usepackage{amsmath}

\usepackage{amssymb}
\usepackage{braket}
\usepackage{color}
\usepackage{fullpage}

\usepackage{mathtools,leftindex,tensor,mhchem}

\usepackage{graphicx}
\graphicspath{{./}{./Figs/}} 

\usepackage{bm}

\usepackage{braket}

\usepackage[colorlinks=true,linktoc=page,linkcolor=blue,citecolor=blue,urlcolor=blue]{hyperref}
\usepackage{cleveref}

\usepackage{epstopdf}

\newcommand{\beq}{\begin{eqnarray}}
\newcommand{\eeq}{\end{eqnarray}}
\newcommand{\bea}{\begin{eqnarray}}
\newcommand{\eea}{\end{eqnarray}}

\def\<{\langle}
\def\>{\rangle}

\newcommand{\hide}[1]{}  

\usepackage{tikz}
\usetikzlibrary{quantikz2}

\newcommand{\sectA}[1]
{
	\addtocounter{subsubsection}{1}
	\setcounter{subsection}{0}
	\ \\
	\pdfbookmark[1]{\thesubsubsection. \ #1}{sect.\thesubsubsection}
	{\Large\bf $=\!=\!=\!=\!=\!=\;$ [\thesubsubsection] \ #1}
	\nopagebreak
	\vspace*{3mm}
}
\renewcommand{\subsubsection}{\sectA}

\begin{document} 
\title{
Chaos-Mediated Quantum State Discrimination Near Unit Fidelity
}

\newcommand*{\affaddr}[1]{#1} 
\newcommand*{\affmark}[1][*]{\textsuperscript{#1}}

\def\par{\futurelet\next\dopar}
\def\dopar{\let\doparA=\endgraf
           \ifx\next\par \let\doparA\relax \fi
           \ifx\next\unpar \let\doparA\relax \fi
           \doparA}
\def\unpar{\donothing{unpar}} 
\def\donothing#1{}

\author{Sourav Paul \affmark[1] }
\email{souravpl2012@gmail.com; sp20rs034@iiserkol.ac.in}
\author{Anant Vijay Varma \affmark[2]}
\email{anantvijay.cct@gmail.com}
\author{Yogesh N. Joglekar \affmark[3]}
\email{yojoglek@iupui.edu }
\author{Sourin Das \affmark[1]}
\email{sourin@iiserkol.ac.in; sdas.du@gmail.com }

\affiliation{
\affaddr{\affmark[1] Indian Institute of Science Education and Research Kolkata, Mohanpur, Nadia 741246, West Bengal, India.}\\
\affaddr{\affmark[2] Department of Physics, Ben-Gurion University of the Negev, Beer-Sheva 84105, Israel}\\
\affaddr{\affmark[3] Department of Physics, IU Indianapolis, LD 154,
402 N Blackford Street, Indianapolis IN 46202-3217, USA}}

\begin{abstract}

We investigate a “quantum microscope” for qubits based on nonlinear discrete-time chaotic dynamics, which exponentially amplifies the initially small fidelity of a pair of states to a large saturation value ($\sim 1/2$), thereby pushing the Helstrom bound to more accessible values. We show that Bell-type temporal correlations can capture even the minutest differences between two initial states, thus enabling their distinguishability. The cost of distinguishability is quantified in terms of the characteristic waiting time of the evolution, defined as the time after which the temporal correlation of a given initial state begins to diverge exponentially from that of a nearby state. The closer the two states are, the longer this waiting time becomes. By combining chaos with Bell-type temporal correlations, this approach opens unexplored avenues for pushing the limits of precision in quantum metrology.
\end{abstract}

\maketitle

{\textit{\underline{Introduction:-}}}
Quantum state discrimination \cite{Helstrom1969,QSdis2,Chefles01112000,QSdis3,Barnett:09,QSdis1,PT_QSdis} lies at the heart of quantum information processing, with critical implications for quantum computing, communication, and metrology \citep{Giovannetti2011,Toth_2014,RevModPhys.90.035005,PhysRevLett.96.010401,book_1}. While standard techniques such as quantum tomography \cite{QST_1,QST_2,QST_3,QST_4,QST_5,QST_6} and optimized measurement strategies, can distinguish states with sufficiently distinct overlaps, they fail dramatically when the fidelity between states approaches unity. The fundamental barrier is the Helstrom bound \cite{Helstrom1969,Barnett:09}: for states $|\Psi\rangle$ and $|\Phi\rangle$ with near-identical fidelity ($|\langle\Psi|\Phi\rangle|^{2}\to 1$), the minimum error probability saturates to $1/2$, rendering discrimination impossible under conventional approaches. Overcoming this limit demands a radical departure from linear quantum dynamics.\\
%
%
%
In this letter we realize a ``{quantum microscope"} \cite{PhysRevA.62.012307}, which allows distinction of any typical pair of pure states (PoS) of a qubit with arbitrary closeness with reasonable accuracy. We use non-linear fractional conformal (FNLC) maps iteratively to evolve a PoS on the Bloch sphere, implemented using an ancilla driven protocol \citep{Gilyén2016} and exploit the chaotic nature of the map \cite{Kiss2006,PhysRevLett.107.100501,PhysRevA.109.042410,PhysRevA.74.040301,fractal_1,fractal_2} for high resolution. The states in a PoS can be discriminated, exploiting either \textit{(A)} Fatau set or \textit{(B)} Julia set. A significant limitation of existing methods using the Fatou set is that they only apply to strongly restricted set of PoS, specifically those where the two states of the PoS converge to two distinct fixed points under repetitive application of the map \cite{PhysRevA.95.023828}. Our approach overcomes this fundamental limitation. By leveraging the chaotic dynamics inherent to the Julia set, we provide a general framework that is effective for any PoS on the Bloch sphere. We chose correlation coefficient \cite{freedman2007statistics} ($r_{XY}$ in Eqn.[\ref{E04}]) between the observed values $X$ and $Y$ of measured quantities for each of the state in the PoS to characterize the microscope.
We find that after finite iterations $r_{XY}$ saturates to zero implying the ``{chaotic}" nature of the dynamics independent of the closeness of a typical PoS. It is worth mentioning that measuring fidelity ($F$) after iterative evolution of the PoS, is inefficient in characterizing the microscope, as the random nature of the dynamics restricts the average fidelity  to
 $\overline{F} = \overline{|\braket{\Psi|\Phi}|^{2}}= 1/2^{N}$ \cite{chalker} (where $N=1$ for a qubit and states $\ket{\Psi}$ and $\ket{\Phi}$ is the PoS considered.) Nevertheless, $\overline{F}=1/2$ implies randomness (see SM \cite{SM} for plots). Also, assuming preparation probabilities are half each, the Helstrom bound reduces to $\frac{1}{2} (1- \frac{1}{\sqrt{2}}) $  $i.e. $ correct prediction with $ \sim   85 \% $ confidence level.


\par We consider combination of two-time correlations (TTC) $C_{ij}$s similar to Bell-type correaltions \cite{LGI_1,Emary_2014,doi:10.1073/pnas.1005774108,Knee2012,PhysRevLett.107.130402} in order to calculate the measure $r_{XY}$. For a dichotomic ($i.e.$ eigenvalues $\pm1$ only) observable $\hat{Q}$ the  TTC $C_{ij}$, is defined as : $C_{ij} =\frac{1}{2} \braket{ \{ \hat{Q}(t_i), \hat{Q}(t_j) \} }$, where  $\{ \}$ is anti-commutator operation and $\hat{Q}(t_{l})$ is the time evolved operator in the Heisenberg picture at two different times $t=0 <t_{i}<t_{j}$. We find that calculating $r_{XY}$ with even one TTC is enough to capture the ``chaotic" nature of the dynamics and discriminate a PoS typically. Infact, for a suitable choice of measurement operator $\hat{Q}$, which is dictated by the roots of the FLNC map (see SM-\cite{SM}) $C_{ij} \approx \braket{\hat{Q}(t_{i})}$. However, here we consider three-time measurement based Leggett-Garg (LG) parameter   $K_3 = C_{12}+C_{23}-C_{13}$ \cite{LGI_1,Emary_2014,doi:10.1073/pnas.1005774108,Knee2012,PhysRevLett.107.130402} as the ``{quantity}" to calculate $r_{XY}$. The reason is multi-fold: (a) LG parameter not only captures the required randomness to discriminate the states and reduces to $K_{3}= C_{23} \approx \braket{\hat{Q}(t_{2})}$ for appropriate choice of measurement operator $\hat{Q}$ but also (b) violation of LG inequality $i.e.$ $K_{3}>1$ guarantees a ``{faulty measurement device}". This allows self-testing of the microscope's performance in a device-independent manner. (c) Finally, for any given single PoS the difference between the $K_{3}$ parameters of each of the states allows to distinguish them in an experiment. \\

{\textit{\underline{FNLC maps and qubit dynamics:-}}} Extended complex plane can be associated with the Bloch sphere via the stereographic projection, allowing for a projection \( S: \mathcal{H} \longrightarrow \tilde{C} \) from the two-dimensional projective Hilbert space \( \mathcal{H} \) to the extended complex plane \( \tilde{C} \) \cite{Lee2002}.
In the context of the Bloch sphere, a pure qubit state is represented as \( \ket{\psi} = (\zeta_1, \zeta_2) = N (z, 1)^T \), where \( N = 1 / \sqrt{\vert z \vert^2 + 1} \) and the corresponding point on the extended complex plane is given by  $z = \zeta_1 / \zeta_2 = \cot{\frac{\theta}{2}}e^{i \phi}$ (ignoring the global phase) \cite{Lee2002}.
Any mathematical mapping $f(z)$ of the complex number $z$, can be projected back on the Bloch sphere. This projection can be thought of as a discrete time evolution , where the transformation \( z \mapsto f(z) \)  results in an  ``{evolved state}" \cite{Paul_2024}:

\begin{equation}
\vert \tilde{\psi} \rangle= \frac{1}{\sqrt{\vert f(z) \vert^2 + 1}} \  \begin{pmatrix}
f(z) \\ 1
\end{pmatrix}
\label{E01}
\end{equation}

This expression capture how the qubit state evolves through the FNLC map over discrete time $i.e.$ iteration. The following relation schematically shows the direct correspondence between the Bloch sphere state and a point on the extended complex plane: $\ket{\psi} \longleftrightarrow z \mapsto z' = f(z) \longleftrightarrow \vert \tilde{\psi} \rangle$. Since the map $f(z)$ can arbitrarily be defined, we consider FNLC maps of minimal order $i.e.$ FNLC map of second order: $f(z) = p(z)/q(z)$, where $p(z)= a z^2 + b; q(z) =c z^2 + d$. In particular we are interested in: 

\begin{equation}
f(z) = \frac{z^2 + s}{s z^2 + 1}
\label{E02}
\end{equation}

This map under repeated iterations behaves in ``regular" fashion in the Fatou set and in ``{chaotic}" fashion in the Julia set. Parameter ``$s$'' controls the size of the Julia set of the map in Eq.(\ref{E02}) for the qubit dynamics. \\

\begin{figure*}
  \centering 
\includegraphics[width=0.6 \columnwidth]
{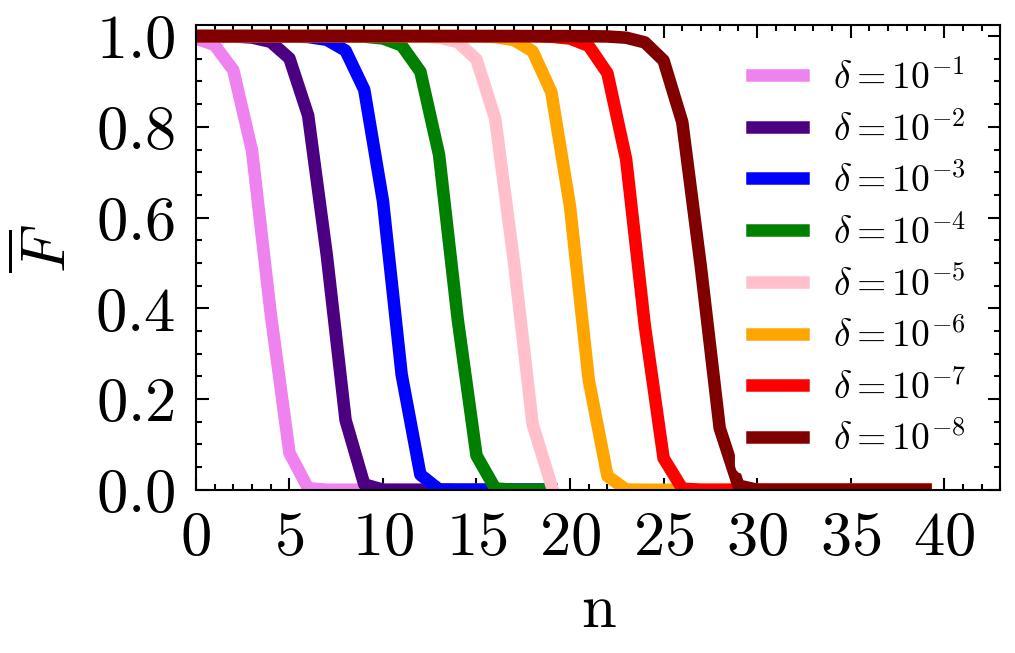}
\includegraphics[width= 0.75\columnwidth]{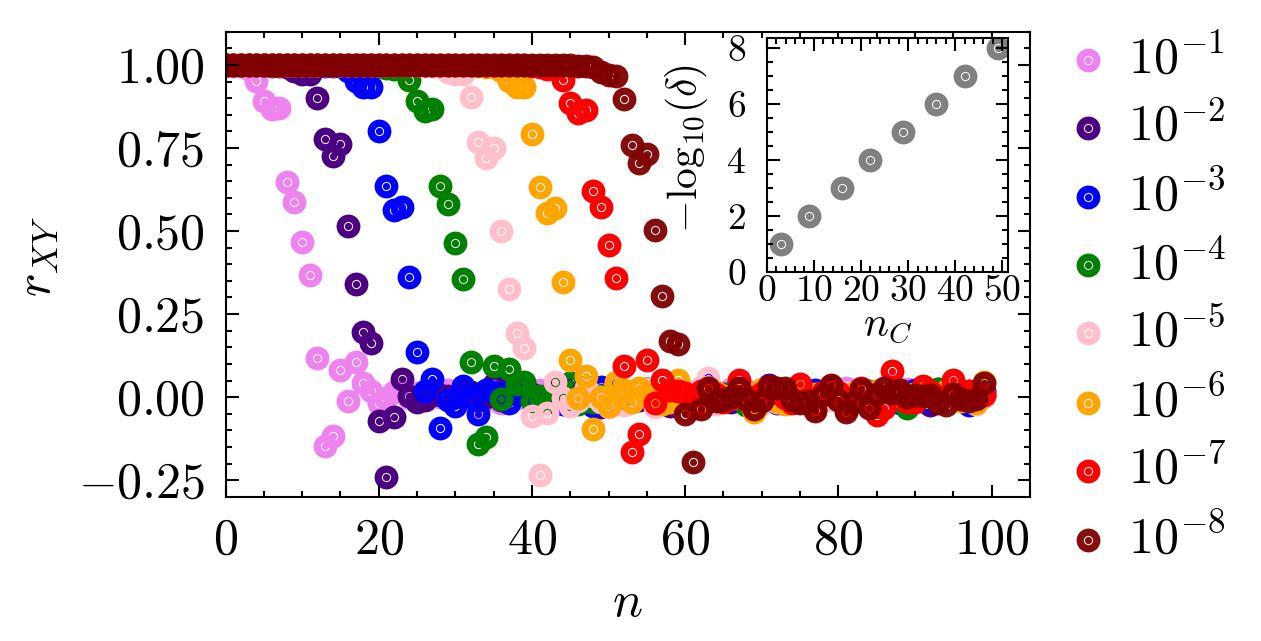} 
\includegraphics[width= 0.7 \columnwidth]{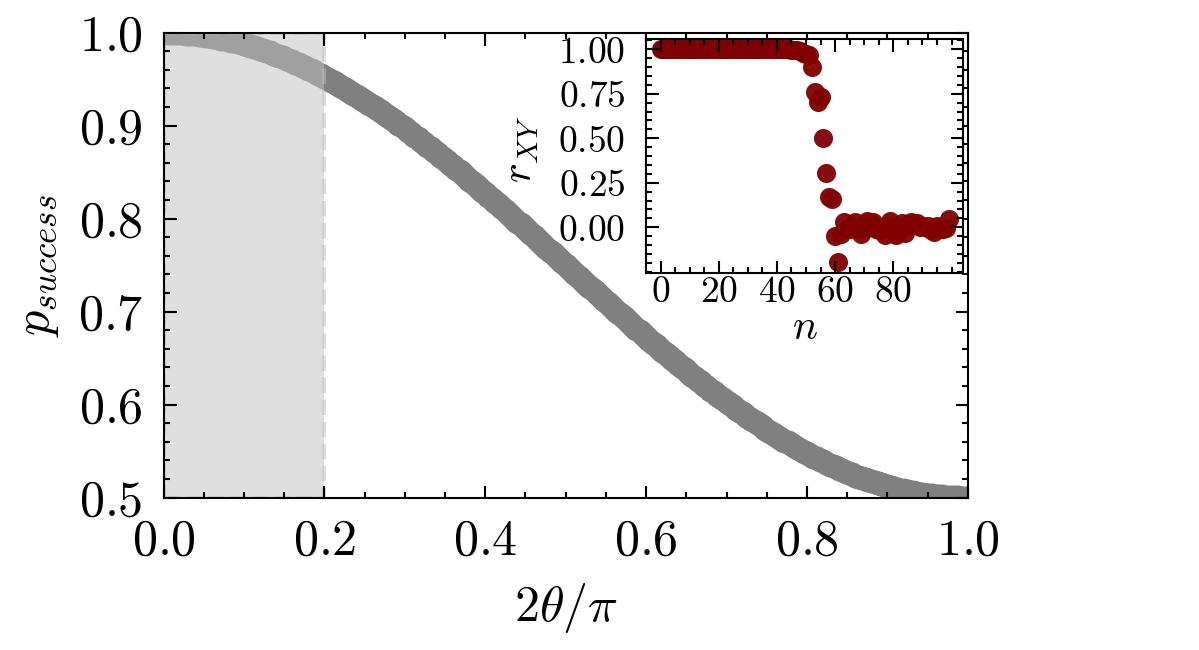}
  \caption{\textbf{(a) Fatau set based discrimination:} Average fidelity of the ensemble of PoS vs the number of iterations. Two states of every PoS in the ensemble are chosen on the either side of the equator and at a distance $\delta$ apart. \textbf{ (b) Julia set based discrimination:} Statistical correlation ($r_{XY}$) vs iteration ($n$).  Here $\delta = 10^{-8}, 10^{-7},...,10^{-1} $ are different orders of separation in the initial PoS which are taken from the whole Bloch sphere for numerical calculation. The FNLC map corresponds to $s=i$ throughout the paper. For numerical calculations, ensemble of  $10^{4}$ complex points (uniformly distributed on the whole Bloch sphere) and their respective pair with separation $\delta$ are chosen. We have chosen the dichotomic observable to be Pauli $\sigma_{x}$ operator from here onwards unless mentioned. \textbf{Inset}: Critical number of iterations needed for discrimination vs the order of separation between the PoS. \textbf{ (c) Optimization of the state discrimination protocol:} Success probability in Eq. (\ref{E05})  wrt $\theta$ (colatitude). \textbf{Inset}: $r_{XY}$ vs $n$ for the initial PoS (chosen from the north pole region belt (grey) with $\theta = 0$ \ \text{to} \  $\theta = \pi/10 $, which are $\delta = 10^{-8}$ distance apart. For numerical calculations, same ensemble size and the measurement operator have been used as in Fig. \ref{fig1}. }
  \label{fig1}
\end{figure*}

{\textit{\underline{(A) Orthogonal fixed points strategy:-}}} For $s=0$ both the north and the south poles are fixed points and are part of Fatau set which comprises all points on the Bloch sphere except the equator. It is known that \cite{PhysRevA.95.023828,Gilyén2016}  any PoS, where the states are separated by the equator eventually evolve to become orthogonal to each other. For the map $f(z)_{s=0}= z^{2},$ we calculate average fidelity for the set of PoS chosen across the equator. The results are shown in Fig \ref{fig1}(a) for different values of $\delta$ \footnote{If the two states of the PoS in the complex plane are $z_{1}$ and $z_{2}$, then those complex points translate to $(\theta_1,\phi_1)$ point (with $z_1 = \cot{\frac{\theta_1}{2}} e^{-i \phi_1}$) and $(\theta_2,\phi_2)$ point (with $z_2 = \cot{\frac{\theta_2}{2}} e^{-i \phi_2}$) in the unit Bloch sphere via stereographic projection. Thereafter we define the $\delta$ as $\delta = \cos^{-1} (\cos\theta_1\cos\theta_2 +\sin\theta_1\sin\theta_2 \cos(\phi_1-\phi_2)) $. For all calculations regarding choosing nearby PoS, we took $\phi_1 = \phi_2$ (same co-latitude points). Moreover, stereographic projection of the states on the Bloch sphere only overestimates the critical number of iterations needed for distinguishability. Therefore, we only consider half of the Bloch sphere in Fig \ref {fig4a} .
   }.  \\
   
{\textit{\underline{(B) Julia set based strategy:-}}} 
 For $s=i$ the whole Bloch sphere corresponds to Julia set. The corresponding dynamics in the sense mentioned above has a stable fixed point determined by $f(z)=z \Rightarrow z=1$ (see SM \cite{SM} for details). Just as the resolving power of a microscope depends on the wavelength of light, the minimum number of iterations needed to resolve any PoS depends upon the location of the PoS w.r.t. the stable fixed point  \cite{Gilyén2016}. \\

{\textit{\underline{Characterization of the Microscope:-}}}  We now present the statistical features of the microscope and issues related to optimization and measurement device induced errors. We begin the outline with the use of the statistical correlation $r_{XY}$ between the LG parameters  ($K_3$s) of ensemble of PoS for the characterization of the microscope. Thereafter we discuss various features of the microscope and discrimination of a given PoS. LG parameter draws a boundary between the classical and the  quantum correlations  that a probability distribution retains as it evolves in time. In a minimal scenario, it requires at least three time instants to distinguish such correlations and this distinction is written in terms of inequality: $K_3 = C_{12}+C_{23}-C_{13},\leq 1$,  where $C_{ij}$s defined as  $C_{ij} =\frac{1}{2} \braket{ \{ \hat{Q}(t_i), \hat{Q}(t_j) \} }$ are two-time correlations defined for a given choice of measurement operator $\hat{Q}= \hat{n} \cdot \vec{\sigma}$, where $\hat{n} = (\sin \theta_{m} \cos \phi_{m},\sin \theta_{m} \sin \phi_{m},\cos \theta_{m})$ can be equivalently written as:


\begin{equation}
C_{ij} = \sum_{\hat{Q}(t_i/t_j) = \pm 1} \hat{Q}(t_i) \hat{Q}(t_j) P_{ij} ( \hat{Q}(t_i), \hat{Q}(t_j)), 
\label{E03}
\end{equation}

where $P_{ij}$s (see SM \cite{SM}) are the joint probabilities of observing outcome $\hat{Q}(t_i)$ and $\hat{Q}(t_j)$ at time instants $t_{i}$ and $t_{j}$ respectively. For this study, we discretize time and focus on the equispaced time interval with $t_1 = 0, \  t_2 -t_1 = n = t_3-t_2$. We must emphasize that $C_{ij}$'s are functions of $n$ which in turn makes $K_3$ a function of $n$ as well. It should also be noticed that for a given initial state $\ket{\psi}$ the time evolved state $e^{-i H (t_j-t_i)} \vert \psi \rangle$ has equivalent form $(f^{(t_j - t_i)}_{ij}(z_{\vert \psi \rangle}),1)^{T}$ with $z_{\vert \psi \rangle}$ being the correspondent complex point to the state $\vert \psi \rangle$ after stereographic projection and where $f^{n}_{ij}(z) = f_{ij} \circ f_{ij} \circ ... \circ f_{ij}(z)$ a composition of map $f_{ij}$ (see SM \cite{SM} for details). We choose an ensemble of PoS as initial states, wherein the states of every pair has \textit{separation} $\delta$ in the complex plane. We then calculate $r_{XY}$ after every iteration, which for two data set $X = \{ x_1, x_2, ..., x_L \}$ and $Y = \{ y_1, y_2, ..., y_L \}$ of population size $L$ is given by:



\begin{equation}
r_{XY} = \dfrac{L \sum_i x_i y_i - (\sum_i x_i)(\sum_i y_i)}{\sqrt{[ L \sum_i x_i^2 - (\sum_i x_i)^2 ] [ L \sum_i y_i^2 - (\sum_i y_i)^2 ]}}
\label{E04}
\end{equation}

\begin{figure*}[htb!]
    \centering
\includegraphics[width=15cm]{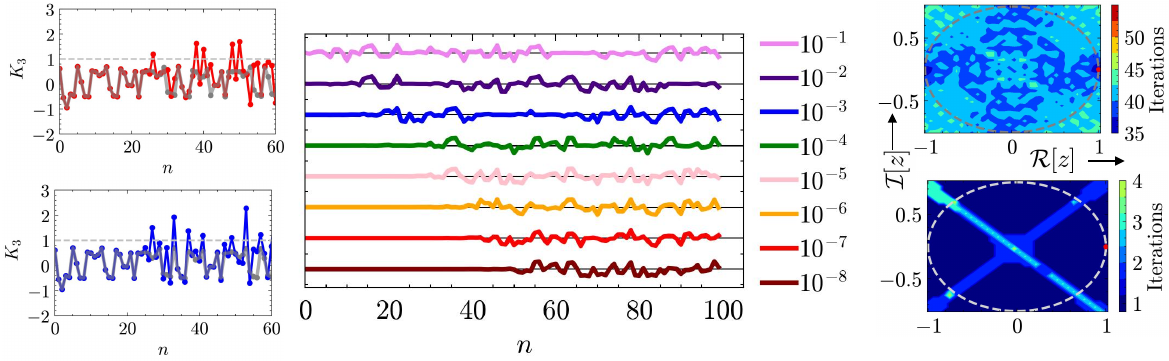}
\caption{(\textit{Left}) \textbf{Sensitivity to the faulty measuring device:} Variation of LG Parameter $K_3$ wrt iteration for randomly chosen initial states. In both the plots gray points represent evolution interrupted with an ideal measurement device, while red markers in (a) shows $K_3$ time series with an error of order $10^{-8} $ in azimuthal angle $\phi_m$  \& (b) same order of error in longitudinal angle $\theta_m$ (blue) of the measurement operator. (\textit{Middle}) \textbf{Difference of $K_3s$ wrt iteration ($n$)}, for an initial PoS chosen randomly from the Bloch sphere. Fluctuation in each curve represents non-zero value of the difference of $K_{3}$s for the chosen PoS. Different colors are the initial separation $\delta$ (of the form $10^{-p}$) between the states. Each curve is bounded from $-2$ to $2$ (y-axis) with the flat lines marking zero value for all cases. (\textit{Right}) \textbf{Resolution heat map:} Discrimination of different PoS in the complex plane, using the difference of $K_{3}$ values. Each point in the plane represents a PoS with separation (upper) $\delta=10^{-1}$ and (lower) $\delta=10^{-8}$. Color is the minimum number of iterations required for non-zero difference of $K_{3}$ values. In both the plots dashed circle marks the boundary of southern hemisphere and red dot indicates the stable fixed point of the dynamics. }
 \label{FIG2}
\end{figure*}

Here, $X$ and $Y$ are LG parameters $K_{3}$ for two states from each pair of the ensemble. The outcome is demonstrated in the Fig \ref{fig1}(b), and the findings are:\ \textit{i}) Correlation $r_{XY}$ is able to capture the ``\textit{chaotic}"  behavior of the discrete dynamics induced by FNLC map since in the large iteration limit $r_{XY}$ stabilizes to $0$. Although, it takes finitely many number of iterations depending upon $\delta$ before the states become distinguishable (not necessarily orthogonal). The randomness ($i.e$  $r_{XY}= 0$) in $K_3$ ensembles of PoS ensures that initial PoS is eventually separated on a measurable scale. \textit{ii}) There is a hierarchy of cost for the magnification power (critical number of iterations) required, which grows linearly with the order of separation of initial PoS implying maximum exponential decay in the probability with $n$ on an average \cite{Gilyén2016}.

\underline{\textit{Success probability of the dynamics:-}}
The success probability of outcome after one iteration ($\ket{\psi} \rightarrow \ket{\tilde{\psi}}$) on one of the qubits for $s= i$ is

\begin{equation}
\ p_{\text{success}}= \dfrac{1+|z|^{4}}{(1+ |z|^{2})^{2}} 
\label{E05},
\end{equation}

where $\dfrac{1}{2} \leq  \ p_{\text{success}} \leq 1$ as $0 \leq |z|  \leq \infty$ (see SM\cite{SM}) shown in Fig.\ref{fig1}(c). Also, the probability of success averaged over full Bloch sphere, after every iteration is  $\left<p_{\text{success}}
 \right>= 2/3$. Therefore, the average probability of the success of the dynamics after $n^{th}$ iteration is roughly $(2/3)^{n}$. Nevertheless, the success probability of the dynamics after $n$ iterations has lower bound of $(1/2)^n$, which is negligible for small separations and requires optimization. \\

\textit{\underline{Optimization of the success probability:-}} In Fig.\ref{fig1} ~(c) and from Eq.(\ref{E05}) we note that the success rate of the dynamics is considerably large around the poles of Bloch sphere. We now chose patches around these poles to define the ensemble of initial PoS, shown as grey shaded region (only for north pole belt shown) in the Fig. \ref{fig1}(c). Calculation of $r_{XY}$ for this patch with the separation between PoS being $10^{-8}$ is illustrated in the Fig. \ref{fig1}(c) (Inset). Note that for south pole patch the inset figure yields same qualitative graph. We note that though $r_{XY}$ is a statistical {quantity}, it works as a good measure of discrimination for small patches as well, as long as the ensemble size is enough. The least value of probability in the shaded regions is $\approx 0.95$ and about $n=60$ iterations (considering worst case) are enough for the correlations to saturate to zero. This implies that even the microscopic distance of the order of $\delta =10^{-8}$ is distinguishable by this protocol successfully with the probability $> \approx 0.05$. Therefore, {apriori} information of the whereabouts of the PoS can drastically improve the success probability of the dynamics. \\

\textit{\underline{Faulty Measurement Device:-}}
We now argue how a ``{faulty measuring device}" can be detected if we use LG parameter based correlation instead of measuring the expectation value of $\hat{Q}$ in time.  Recall that $C_{ij} =P_{ij}(\uparrow,\uparrow) -P_{ij}(\uparrow,\downarrow) -P_{ij}(\downarrow,\uparrow) + P_{ij}(\downarrow,\downarrow) $. Also, consider the root $z=1$ for the eqn. $f(z)= z$, which corresponds to $\hat{n} = \hat{i}$ direction. we can now define the measurement operator to be $\hat{Q}= \vec{\sigma} \cdot \hat{n} =  \sigma_{x}$. Let us now assign the probabilities of measuring $\uparrow$ and $\downarrow$ at time instant $t_{i}$ to be $\alpha$ and $\beta$ respectively ($\alpha+\beta=1$). However, once we have measured the states at time instant $t_{i}$ resulting in one of the eigenstates of the $\sigma_{x}$ we know {apriori} the probabilities at time instant $t_{j}$ as the eigenstates of the $\sigma_{x}$ are fixed points of the dynamics. This allows us to write $C_{ij} = \alpha-\beta = \braket{\psi_{t_i}| \hat{Q} |\psi_{t_i}} = \braket{\hat{Q}(t_{i})}$. Since in this protocol we have considered $t_{1}=0;  t_{2}=n$ and $t_{3}= 2n$ and the dynamics cease once measured implies $C_{12}= C_{13}$. Therefore, $-1 \leq K_{3} = C_{23} = \braket{\hat{Q}(t_{2})} = \braket{\hat{Q}(n)} \leq1$ $i.e.$ LG inequality is satisfied, if the measuring device is ideal. However, once we have ``{faulty measurement device}" $i.e.$ error in 
$(\theta_{m},\phi_{m} )$, LGI is  violated {independent} of the choice of PoS. This fact is illustrated in Fig.\ref{FIG2} (left panels) for a randomly selected state on the Bloch sphere. Since the FLNC map discussed here is not a CPTP map, and is the example of post-selected dynamics like non-Hermitian dynamics we expect the LG parameter to even violate Luder's bound \cite{Varma_2021,PhysRevA.108.032202}.

\textit{\underline{Discriminating a given PoS:-}} After discussing the characterization and issues related to errors, we finally illustrate how to distinguish any given PoS. While Fig. \ref{fig1} displays a {typical} scenario, it is essential to show how to distinguish a given PoS in an experiment. For this purpose we calculate LG parameters for each of the states in the PoS after every iteration and the first non-zero value in their difference marks the number of iterations needed to distinguish the states. Note that $K_{3}= \braket{\hat{Q}(t_{2})} = \braket{\sigma_{x}(n)}$. Representative cases for different orders of separation $\delta$ are displayed in the Fig. \ref{FIG2}. As mentioned above the resolution power of the microscope depends upon the location of the PoS wrt fixed point of the dynamics. We demonstrate this statement in the Fig. \ref{FIG2} (\textit{right panels}) for all the PoS selected on the Bloch sphere  (southern hemisphere).

\begin{figure}[htb]\centering
\includegraphics[width= 0.49\columnwidth]
{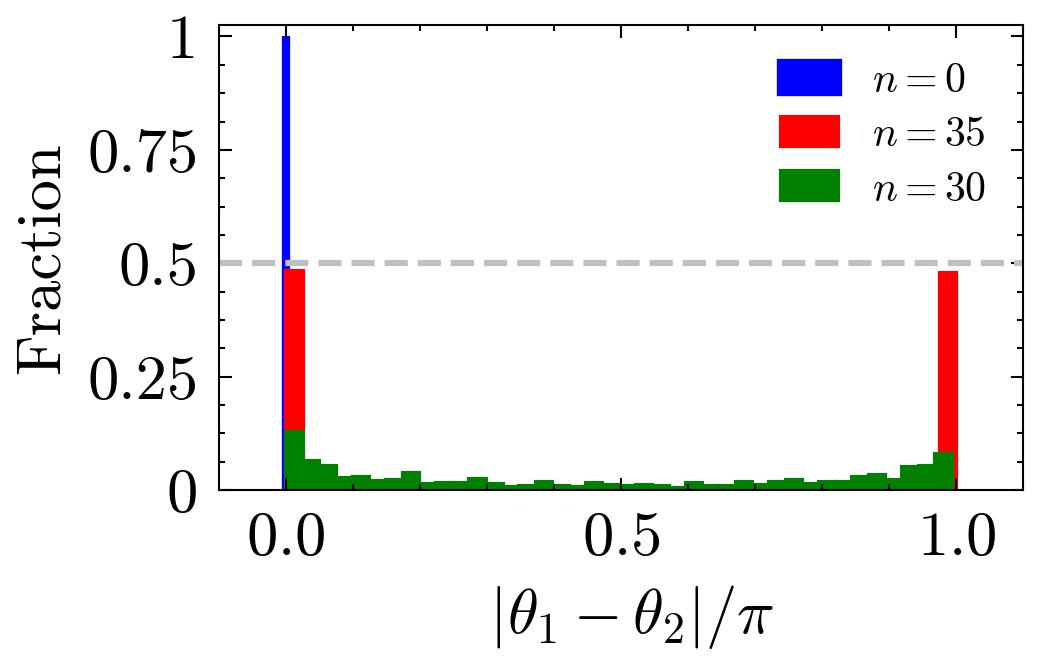}
\includegraphics[width= 0.49\columnwidth]
{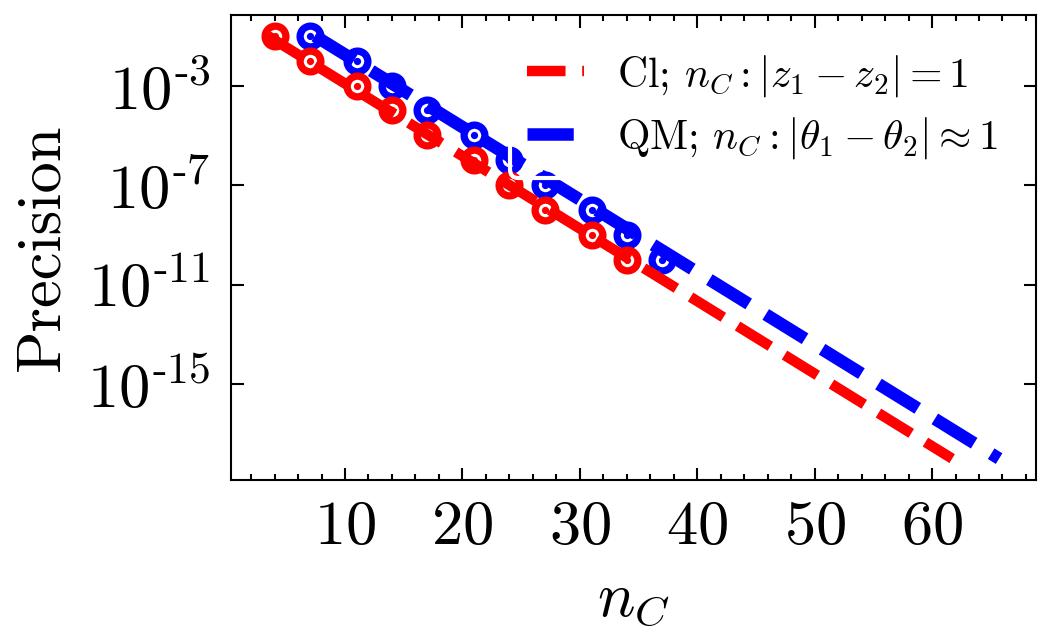}
\caption{\textbf{Efficiency  \& Machine Precision:} (a) Typical histogram (for strategy (A)) showing the fraction of the population of the random pair of points used for the simulation after different iterations, where $|\theta_{1}-\theta_{2}|/\pi =0,1$ implies co-linear and orthogonal states respectively. (b) Stipulated precision vs critical number of iterations required for the precision. Dashed lines are extrapolated curves. Dots are data obtained from simulation. }
 \label{fig7}
\end{figure}

{\textit{\underline{Discussion:-}}} In this section we contrast the strategy (B) with previous works and mention few observations on the limitations and offshoots of our study. In strategy (A)($i.e. \ s=0$)  except for equator rest of the Bloch sphere points belong to Fatou set. Since in this case the two poles are the attractive fixed points of the dynamics. Therefore, states which are separated by equator eventually ends up on the two opposite poles implying orthogonality. The result of the simulation is illustrated in the Fig \ref{fig1} (a). We find that strategy (A) only works if the two states of the PoS are on the either side of the equator, therefore a better approach when this distinction is known beforehand. However, when this distinction is not known apriori, then strategy (B) is more suitable compared to (A). Moreover, the critical number of iterations required in (A) is limited to half of the trials as demonstrated in Fig. \ref{fig7}(a). However, we find as an offshoot of our study that strategy (A) can be utilized as a machine precision tool, for both classical and quantum machines (see Fig. \ref{fig7}(b)). Also in strategy (B), the identification of a suitable measure $i.e.$ the difference of LGI to distinguish the states is the novelty of this study. While other strategies like tomography and feedback control with weak measurement \cite{PhysRevA.62.012307}, explicitly determine the two states by measuring all spin components of the state vector, our strategy only measure one component and answer whether the states are distinct or not without ever identifying the states. This one spin component corresponds to the fixed point of the dynamics and require no further measurements to calculate two-time correlations. Though, this strategy relies on the assumptions that the quantum circuit for implementing an iteration (evolution) is ideal and infinitely large ensemble size is available to exploit in general for measurement statistics.

{\textit{\underline{Summary:-}}} To summarize, we have investigated a possibility of  high resolution ``\textit{quantum microscope}" for a qubit exploiting the fractal nature of discrete emulated dynamics. The highlights of this study are \textit{(i)} the statistical correlation $r_{XY}$ can efficiently characterize this microscope for optimized choice of the measurement operator. This correlation captures the ``chaotic" nature of the discrete dynamics and can be utilized as an alternative to the Lyapunov Exponents. In general, the cost of distinguishing the states increases exponentially with the decrease in the initial distance between them and therefore, exponentially larger size of the initial ensemble is required for higher resolution in this protocol \cite{Gilyén2016}. \textit{(ii)} The difference of LG parameters of the two states in PoS is enough to discriminate them in an experiment. \textit{(iii)} The LG inequality acts as a testbed for detecting a faulty measurement device.\\

{\bf \textit{ Acknowledgment:-} } S.P. offers his gratitude to the Council of Scientific and Industrial Research (CSIR), Govt. of India for financial support. A.V.V. would like to acknowledge the Israel Science Foundation (Grant No.518/22) for funding. \\
S.P. and A.V.V. contributed equally to this work.

\clearpage
\onecolumngrid
\pagestyle{empty}

\begin{center}
{\large \bf Chaos-Mediated Quantum State Discrimination Near Unit Fidelity}

{Sourav Paul, Anant Vijay Varma, Yogesh N. Joglekar, and Sourin Das}

\ \\

{\large (Supplementary Material)}

\end{center}

In this supplementary we provide details of the FLNC maps and its Julia set, calculation of statistical measure $r_{XY}$, success probability of the emulated discrete dynamics, calculation of discrete time LGI, details of quantum circuit  implementation for $n^{th}$ iteration.


\sectA{Fractal nature of the FNLC map $f(z)$ \label{SM1}}

\begin{figure}[htb!]
\centering
\includegraphics[width=0.23 \columnwidth]{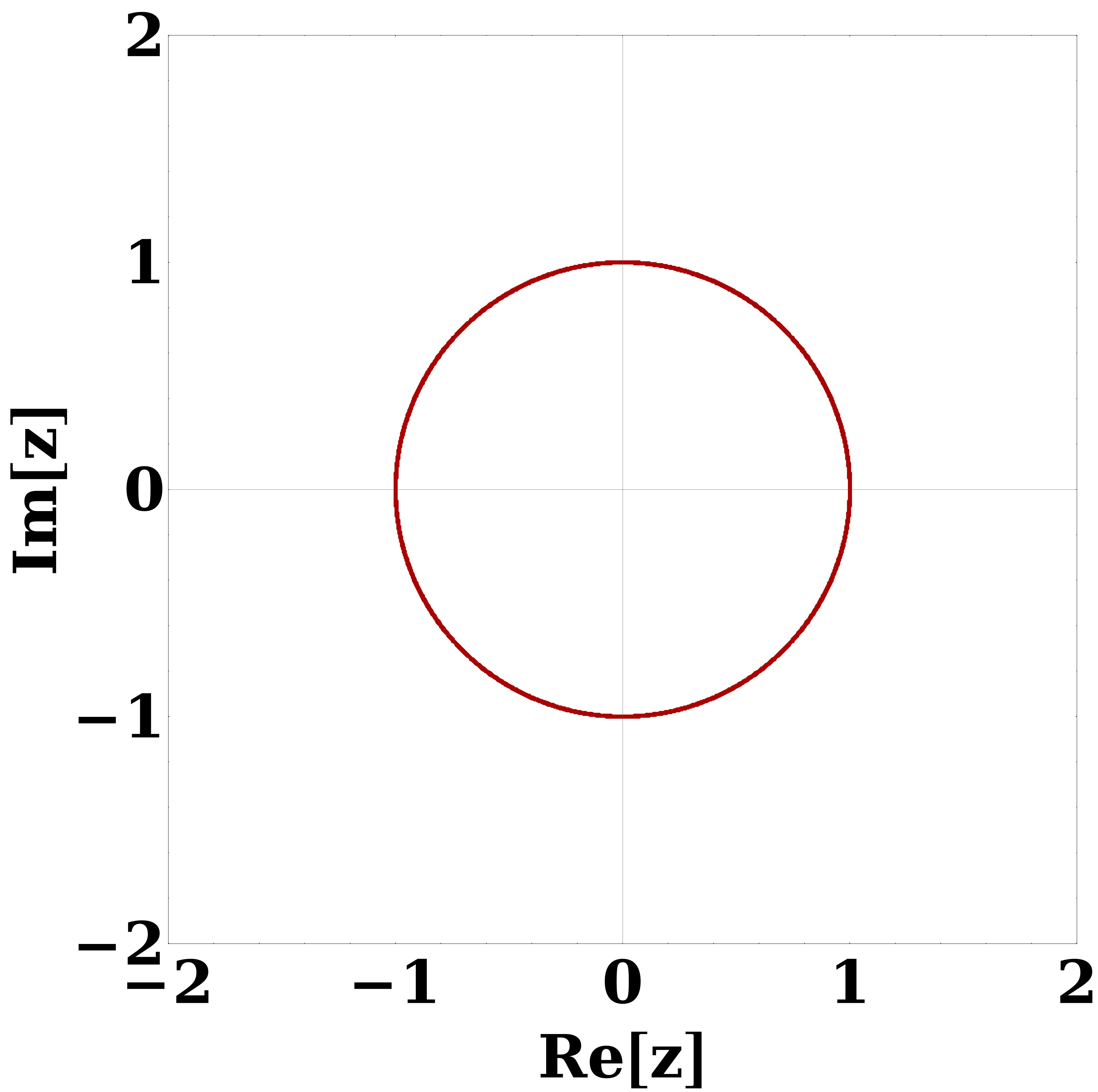} 
\includegraphics[width=0.23 \columnwidth]{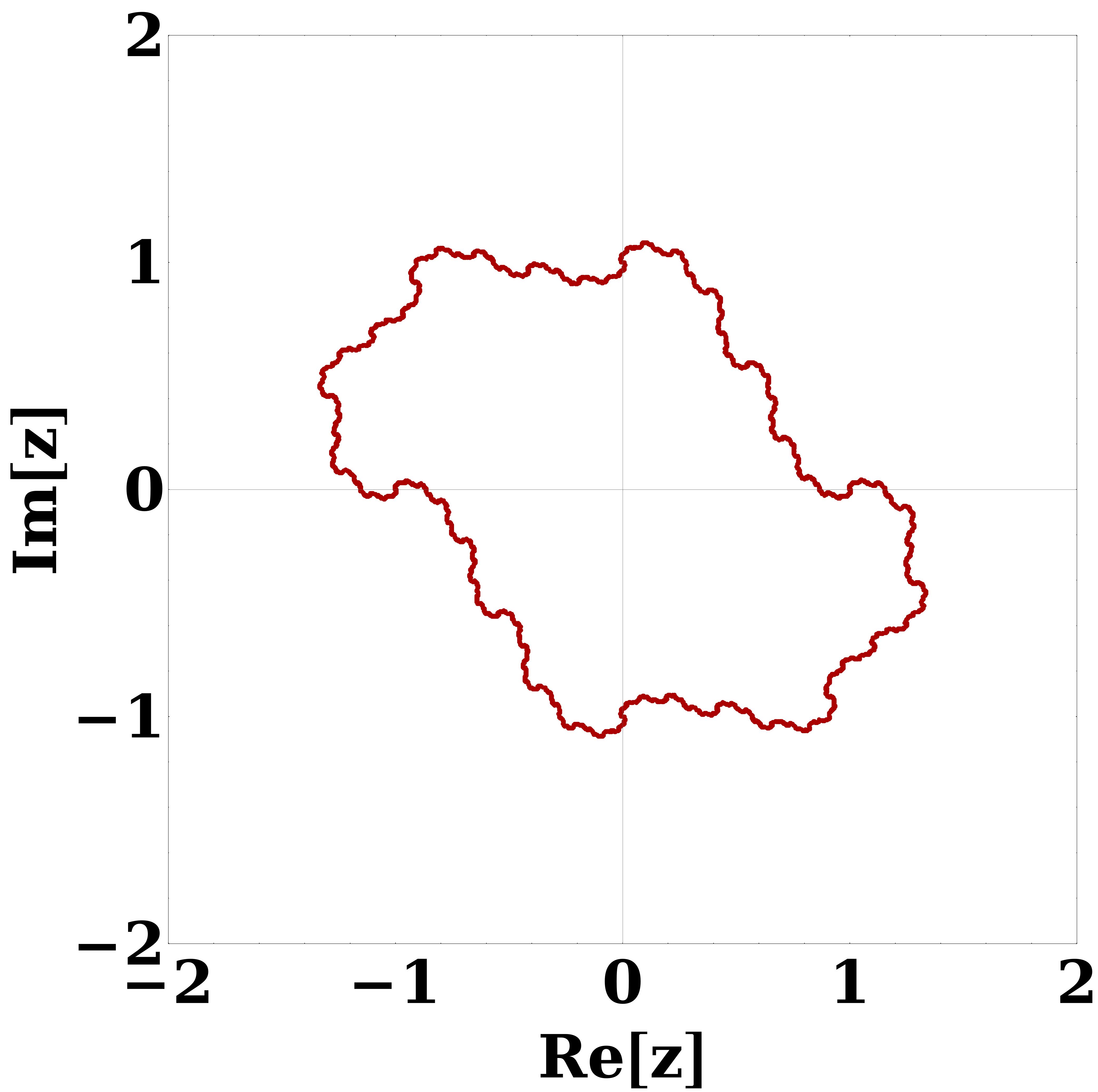} 
\includegraphics[width=0.23 \columnwidth]{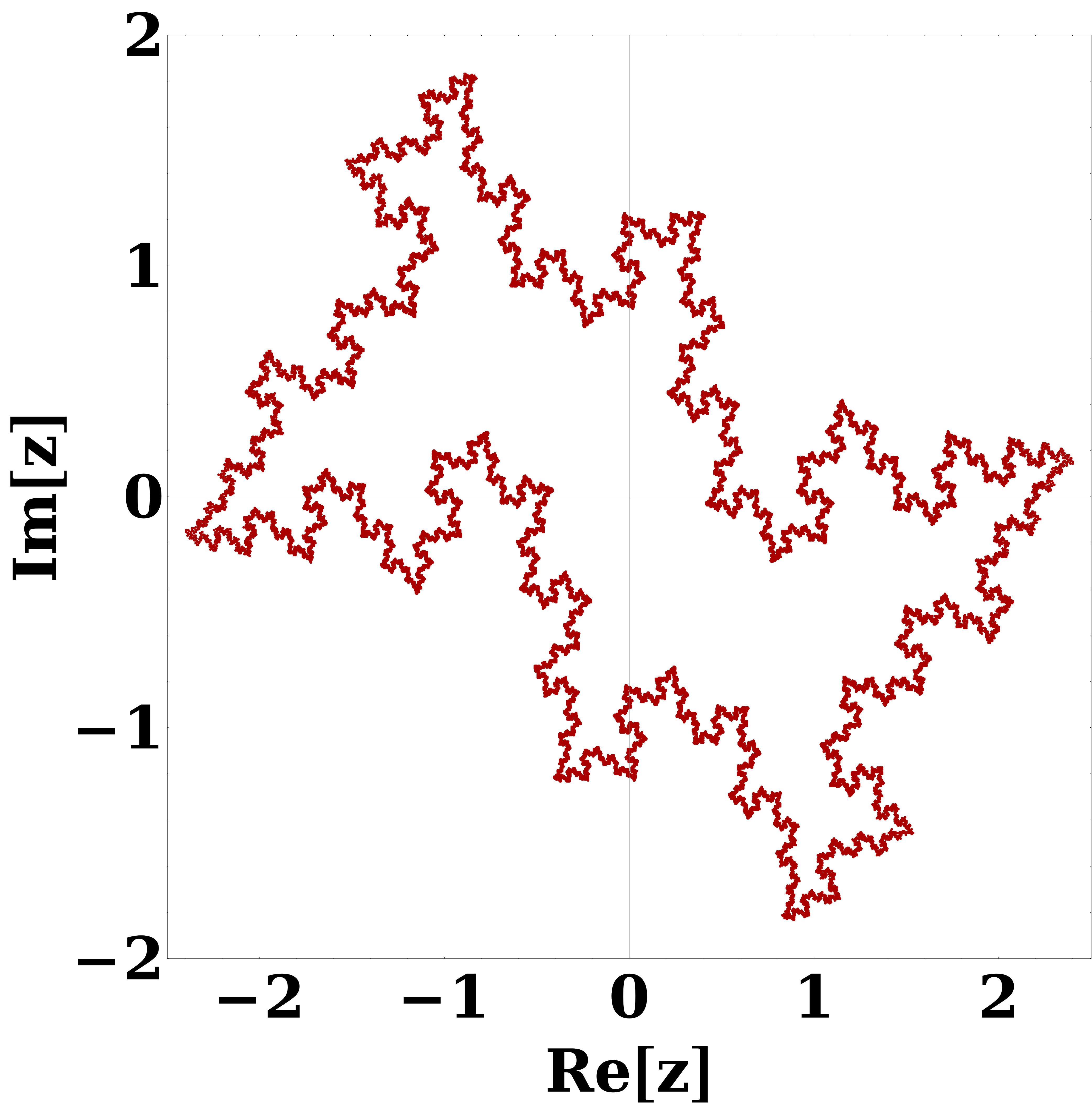} 
\includegraphics[width=0.23 \columnwidth]{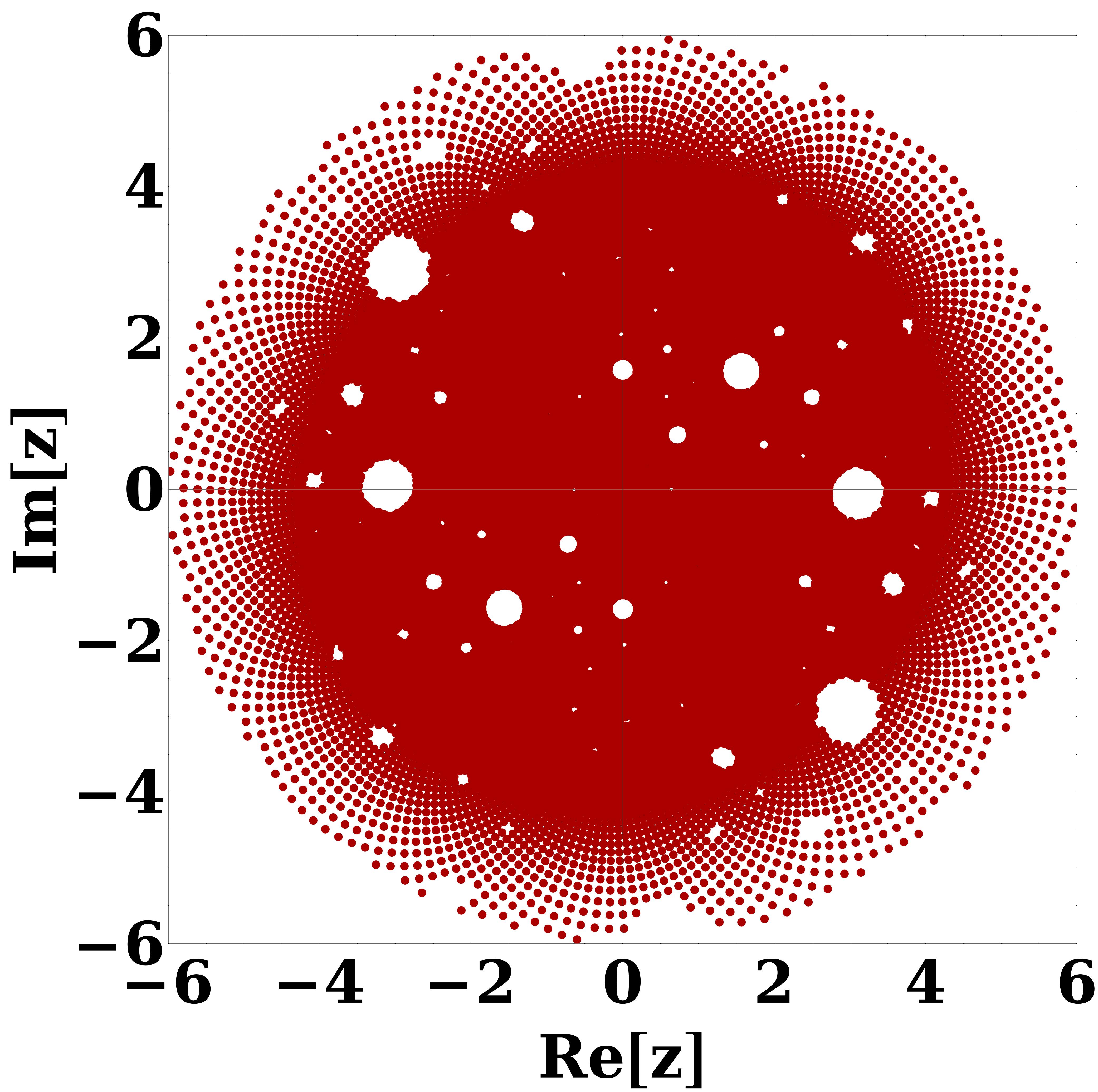} 
\caption{Julia Set plots in complex $z$ plane with increasing value of parameter $s$. Corresponding $s$ values are $0, \ 0.25i, \ 0.5i, \ 0.99i$ from left to right respectively. Here empty regions correspond to Fatau sets in all the plots.}
\label{S1}
\end{figure}

\begin{figure}[htb!]
\centering
\includegraphics[width=0.4 \columnwidth]{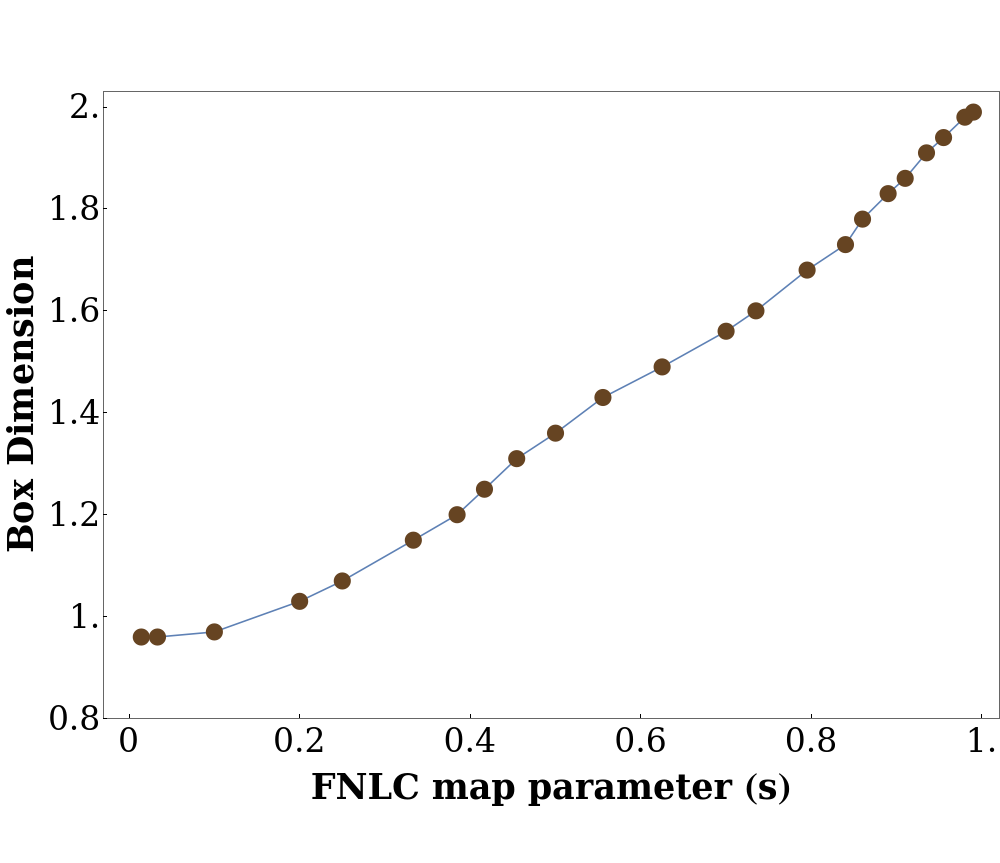} 

\caption{Plot of Box dimension (fractal dimension) as function of $s$ parameter of the FNLC map (Eq.\ref{E02}).} 
\label{S2}
\end{figure}

FNLC maps show fractal nature upon repeated iteration. Below we show Julia set of this FNLC map (given in main text Eq.(\ref{E02})) for different $s$ parameter value (in Fig.\ref{S1}). Behavior of Julia set can be measured using various fractal dimensions \cite{fractal_2}. In the Fig.\ref{S2} we show the Box dimension as a function of control parameter $s$.

\sectA{Discrete Time LGI with FNLC Maps and Evaluation of Statistical Correlation $r_{XY}$ \label{SM2}}

In this supplementary section, we outline the procedure for evaluating the three-time Leggett-Garg (LG) parameter $K_3$ (given in maintext)  within the framework of discrete-time evolution governed by FNLC maps. These maps are described by Eq.(\ref{E02}), with the measurement process characterized by the dichotomic operator $\hat{Q}= \vec{\sigma} \cdot \hat{n}$. The evaluation involves the following systematic steps: \\
\textit{\underline{Steps:}}
(\textit{i}) \ Initial Evolution ($t_1 \longrightarrow t_2$): \ The state of the qubit system at the initial time 
$t_1$ is represented by the complex coordinate $z = z_1$ on the extended complex plane. The evolution from 
$t_1$ to $t_2$ is induced by the map 
$f_{12}(z)$ (the same map used in Eq.(\ref{E02})), which acts on the initial state.
\par (\textit{ii}) \ Subsequent Evolution ($t_2 \longrightarrow t_3$): \ The system then evolves over the next time interval $t_2$ to $t_3$. The state at time $t_2$, denoted by $z = z_2 = f_{12}(z_1)$, undergoes further evolution governed by the map $f_{23}(z)$ (the same map used in Eq.(\ref{E02})).
\par (\textit{iii}) \ Composite Evolution ($t_1 \longrightarrow t_3$): The total evolution from the initial time $t_1$ to the final time 
$t_3$ is described by the composition of the individual maps. Explicitly, the composite map $f_{13} = f_{23} \circ f_{12}$ acts on the initial state $z_1$, thereby encapsulating the overall system dynamics across the three discrete time steps.

\begin{equation*}
z_2 = f_{12}(z_{1}) = \frac{ \ z_{1}^2 + s}{s \ z_{1}^2 + 1}; \ \  z_3 = f_{23}(z_{2}) = \frac{ \ z_{2}^2 + s}{s \ z_{2}^2 + 1}
\end{equation*}
\begin{equation}
z_3 = f_{23}(f_{12}(z_1)) =  \frac{ \ \Big({\frac{\ z_{1}^2 + s}{s \ z_{1}^2 + 1}}\Big)^2 + s}{s \  \Big({\frac{\ z_{1}^2 + s}{s \ z_{1}^2 + 1}}\Big)^2 + 1}
\label{E06}
\end{equation}

\underline{Joint Probabilities ($P_{ij})$:} \ The temporal correlations $C_{ij}$'s (given in main text) are expressed in terms of joint probabilities $P_{ij}$'s which are required to evaluate the LG parameter $K_3$ for the dichotomic observable $\hat{Q} = \vec{\sigma} \cdot \hat{n}$.
\begin{equation}
 \quad  P_{ij} (\hat{Q}(t_i),\hat{Q}(t_j)) =    
\frac{(\vert \langle \pm_Q \vert e^{-i H (t_j-t_i)}\vert \pm_Q \rangle \vert^2) (\vert \langle \pm_Q \vert e^{-i H (t_i-t_1)}\vert \psi^{(0)} \rangle \vert^2 )}{\langle \pm_Q \vert e^{i H^{\dagger} (t_j-t_i)} e^{-i H (t_j-t_i)} \vert \pm_Q \rangle  \langle \pm_Q \vert e^{i H^{\dagger} (t_i-t_1)} e^{-i H (t_i-t_1)} \vert \psi^{(0)} \rangle}
\label{E07}
\end{equation}

where, $i<j \ \{i,j = 1,2,3\}$ and $\ket{\psi^{(0)}}$ is describing initial state. Here $\hat{Q}(t_k)$ denotes the measurement outcome either $+1$ (corresponding to the $\ket{\uparrow}_n $) or $-1$ (corresponding to the $\ket{\downarrow}_n $)  of dichotomic observable $\hat{Q}  = \vec{\sigma} \cdot \hat{n}$. It is to be pointed out that, the state $e^{-i H (t_j-t_i)} \vert \phi \rangle$ is equivalent to the state $(f_{ij}(z_{\vert \phi \rangle}),1)^{T}$ where $z_{\vert \phi \rangle}$ is complex point residing in complex plane referring the state $\vert \phi \rangle$ on the Bloch sphere via stereographic projection. Also $f_{ii}(z_{\vert \phi \rangle}) = z_{\vert \phi \rangle}$ for all $\{i = 1,2,3\}$. The $f_{ij}(z_{\vert \psi \rangle})$ is the same map as described by Eq.(\ref{E06}).\\
\par Using the expression of FNLC map (Eq.(\ref{E06})),  we find out the joint probabilities (Eq.(\ref{E07})) and thereafter the correlation functions ($C_{ij}$) (given in main text Eq.(\ref{E03})), corresponding to a initial state lying on the Bloch sphere undergoing discrete time evolution induced by FNLC maps (of the form Eq.(\ref{E02})) for the dichotomic measurement operator $\hat{Q}= \vec{\sigma} \cdot \hat{n}$ (Eq.(\ref{E06})-Eq.(\ref{E07})).\\ \\

\underline{Generation of $K_3$ data series over iteration for ensemble of states}:\ The above process summarizes the generation of $K_3$ data series over iteration (discrete time) for a specific initial state with a fixed dichotomic observable observable and a specific FNLC map. This process can be replicated for an ensemble of different initial states carefully chosen from the Julia set of a specific FNLC map (Eq.\ref{E02}) lying on the Bloch sphere.\\ \\
\underline{Evaluation of $r_{XY}$}: To evaluate statistical correlation ($r_{XY}$) (in main text Eq.(\ref{E04})) for the ensemble of $K_3$ data series, we first generate the $K_3$ data series for initial and corresponding paired states from the different PoS which are $\delta$ distance apart on the Bloch sphere taken from the Julia set. Next fixing an particular iteration($n$) value, we collect all the $K_3$ data i)  for every initial state for that $n$ and call that data set as $X$, ii) \ for the corresponding pair of the initial states for that $n$ and call that data set $Y$. After that we find the standard statistical correlation ($r_{XY}$) for these two data sets and plot them as a function of $n$ (no. of iteration).\\ \\
\underline{Analytical Form of $r_{XY}$}: \ In this para-section we present an analytical form of $r_{XY}$ as a recursion relation varying with iteration ($n$). Following $K_3^{(n)} = \langle \sigma_x \rangle^{(n)} = \alpha - \beta = \frac{2 \text{Re}(f^{(n)}(z))}{1+\vert f^{(n)} \vert^2} = \sin \theta_n \cos \phi_n,$ we can turn the statistical correlation ($r_{XY}$) (in main text Eq.(\ref{E04})) as integral over $\theta_n$ and $\phi_n$ having a recursion relation over $n$. Let us assume that $X$ corresponds to all $K_3$ data for every initial state on the Bloch sphere whereas $Y$ corresponds to all partner pair $K_3$ data for those particular initial states. This will make $X$ and $Y$ a functions of $\theta_n$, $\phi_n$ and $\tilde{\theta}_n, \tilde{\phi}_n$, or more precisely $X \equiv X(\theta_n,\phi_n)$ and $Y \equiv Y(\tilde{\theta}_n,\tilde{\phi}_n)$. Therefore in integral language, $r_{XY}$ becomes,
\begin{equation}
r_{XY}^{(n)} = \frac{\int \int d\theta_0 d\phi_0 X(\theta_n,\phi_n)Y(\tilde{\theta}_n,\tilde{\phi}_n)}{\sqrt{\int \int d\theta_0 d\phi_0 X^2(\theta_n,\phi_n)} \sqrt{ \int \int d\theta_0 d\phi_0 Y^2(\tilde{\theta}_n,\tilde{\phi}_n)}}
\label{E08}
\end{equation}
where $\delta$ is the initial (zeroth iteration) separation of two states typically $10^{-8},10^{-7},...,10^{-1}$ (As taken in our manuscript), while $\theta_0 = \theta, \phi_0 = \phi, \tilde{\theta}_0 = \theta + \delta, \tilde{\phi}_0 = \phi$ (see [39]). The recursion integral in Eq.(\ref{E08}) can be ultimately reduced to $\theta_0, \phi_0$ (see [39]). For zeroth iteration the $r_{XY}^{(0)}$ integral reduces to,
\begin{equation*}
r_{XY}^{(0)} = \frac{\int \int d\theta_0 d\phi_0 X(\theta_0,\phi_0) Y(\tilde{\theta}_0, \tilde{\phi}_0)}{\sqrt{\int \int d\theta_0 d\phi_0  X^2(\theta_0,\phi_0)} \sqrt{ \int \int d\theta_0 d\phi_0 Y^2(\tilde{\theta}_0,\tilde{\phi}_0)}} = \frac{\int \int d\theta d\phi \cos^2\phi \sin \theta \sin (\theta+\delta)}{\sqrt{ \int \int d\theta d\phi \cos^2\phi \sin^2 \theta} \sqrt{ \int \int d\theta d\phi \cos^2\phi \sin^2 (\theta+\delta)}} = \cos \delta
\end{equation*}
\sectA{Average Fidelity for pairs of initial states \label{SM3}}

In this supplementary section we evaluate average fidelity between PoS, which are separated by $\delta$ distance initially, lying on the Julia set of the Bloch sphere of the FNLC maps. We define the fidelity as,

\begin{equation}
F(n) = \vert \langle \psi_{\delta}(n) \vert \psi(n) \rangle \vert^{2} \quad \text{with} \quad \vert \psi(n) \rangle = \frac{1}{\sqrt{1 + \vert f^{(n)}(z) \vert^2}}\begin{pmatrix}
f^{(n)}(z) \\ 1
\end{pmatrix}, \quad \vert \psi_{\delta}(n) \rangle = \frac{1}{\sqrt{1 + \vert f^{(n)}(z_{\delta}) \vert^2}} \begin{pmatrix}
f^{(n)}(z_{\delta}) \\ 1
\end{pmatrix}
\label{E09}
\end{equation}

where $\vert \psi(n) \rangle$ and $\vert \psi_{\delta}(n) \rangle$ corresponds to state and pair partner of that state separated by distance $\delta$, with $z$ and $z_{\delta}$ being the complex points corresponding to initial state and its pair of the PoS for $n=0$. $n$ denotes the iteration.

\begin{figure}[htb!]
    \centering
\includegraphics[width=0.5\columnwidth]{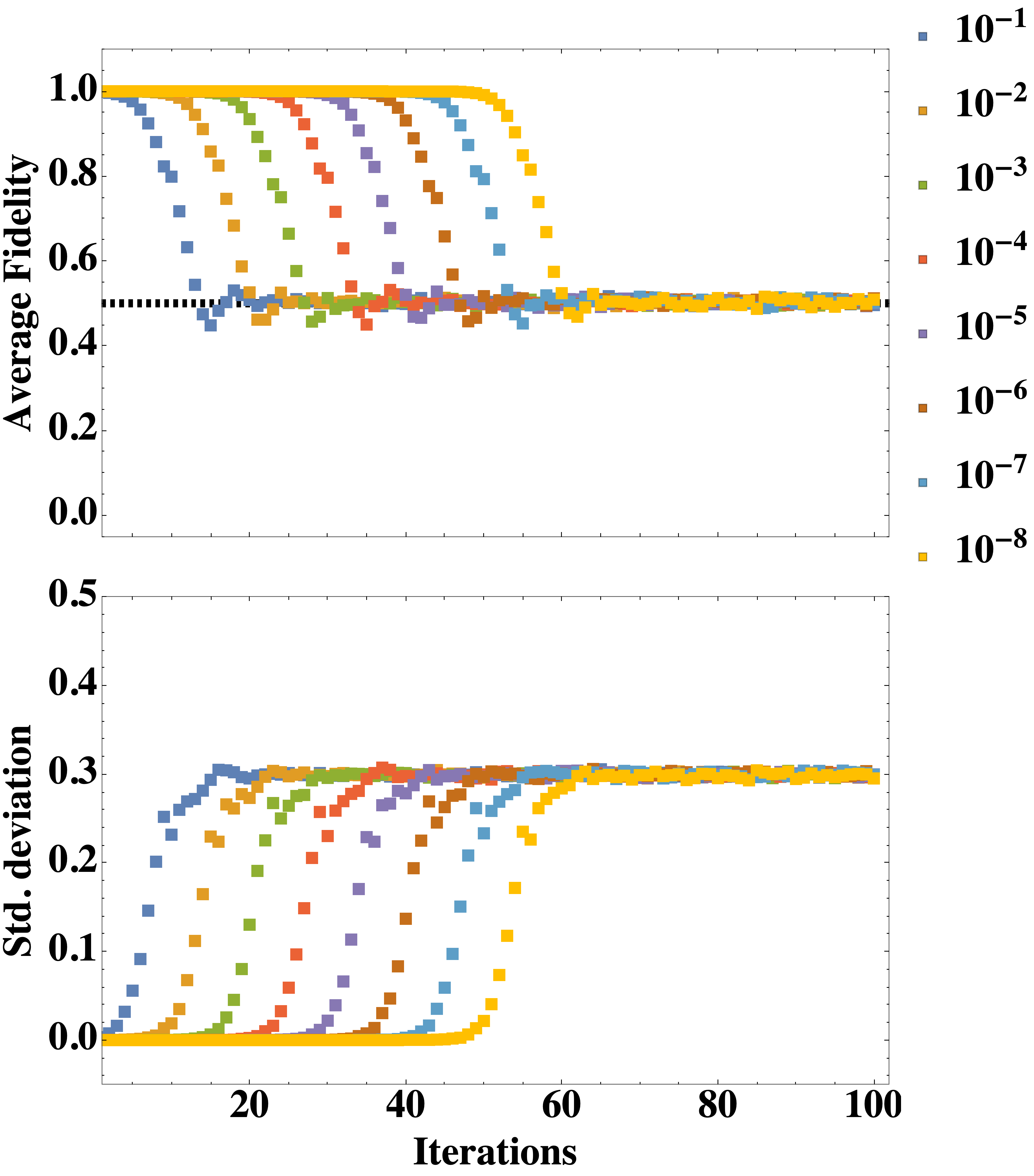} 
\caption{\textbf{Average \& std. deviation of Fidelity vs iterations (n):} Here $\delta = 10^{-1},..., 10^{-8}$ suggests the separation of the initial PoS which are taken from the whole Bloch sphere for numerical calculation.Here the $s=i$. Here we take $10^4$ (=$100 \times 100$) points uniformly distributed over the Julia set of the whole Bloch sphere.}
\label{S3}
\end{figure}
\par Following the same procedure for choosing initial states as discussed in main text, we evolve both the initial states and the correspondent paired states via the FNLC map (Eq.(\ref{E02})) which are chosen from the Julia set of the Bloch sphere and evaluate the fidelity as a function of $n$ (iteration). Next we take the average of the fidelity found for each pair of initial states separated by $\delta$ distance and plot them as function of $n$ along with standard deviation. The outcome is demonstrated in the Fig \ref{S3}, and the findings are: \ \textit{i}) Average fidelity fails to distinguishes a typical PoS, as fidelity saturates to highly non-zero value along with fact that standard deviation around the average fidelity is typically large for all scales of separation of the initial PoS. \textit{ii}) It is although interesting to note that average fidelity guides in detecting the critical iteration to look for to perform the statistical correlation measure.\\ \\


\sectA{Ancilla based quantum Circuit for generating $n$-th iteration of FNLC map \label{SM4}}


In this supplementary section we elaborately discuss the  implementation of the FNLC map $f(z) = \frac{z^2+s}{s z^2 + 1}$ (where $s$ being purely imaginary varying from $s=0$ to $s = i$) starting from identical two qubit system. We evolve the two qubit system via $4 \times 4$ unitary matrix only to measure the one of the qubits (acting as ancilla), in the $\sigma_z$'s positive eigenstate at the end. This measurement induced state via post-selection generates a non unitary evolution of the first qubit with some finite success rate which one can post-select and renormalize to make it a pure state on the Bloch sphere. Below we give a brief mathematical description of the same. One starts with the initial two qubit seperable state as,
\begin{equation*}
\vert \Psi \rangle = \vert \psi \rangle \otimes \vert \psi \rangle = \frac{1}{\sqrt{1+\vert z \vert^2}} \begin{pmatrix}
z \\ 1
\end{pmatrix} \otimes \frac{1}{\sqrt{1+\vert z \vert^2}} \begin{pmatrix}
z \\ 1
\end{pmatrix} 
\end{equation*}
\begin{equation}
= \frac{1}{{1+\vert z \vert^2}} \begin{pmatrix}
z^2 \\ z \\ z \\ 1 
\end{pmatrix}
\label{E10}
\end{equation}
Next one implements two 2-qubit unitary gates  $U_{XOR}$ and $U_{\text{gate}}$ (FIG.\ref{fig5}) in succession. The matrix form of the gates are
\begin{equation}
U_{XOR} = \begin{pmatrix}
1 & 0 & 0 & 0 \\
0 & 1 & 0 & 0 \\
0 & 0 & 0 & 1\\
0 & 0 & 1 & 0
\end{pmatrix},   U_{\text{gate}} = \frac{1}{\sqrt{1+\vert s \vert^2}} \begin{pmatrix}
1 & 0 & s & 0 \\
0 & s^{*} & 0 & 1 \\
s & 0 & 1 & 0\\
0 & 1 & 0 & s^{*}
\end{pmatrix}
\label{E11}
\end{equation}
For $s=0$ and $s=i$, $U_{\text{gate}}$ can be written in terms of $2 \times 2$ Identity matrix ($\mathcal{I}$) and pauli matrices ($\sigma_i$'s) as,
\begin{equation*}
U_{\text{gate}}(s=0) = \frac{1}{2}\Big(\mathcal{I} \otimes \mathcal{I} + \mathcal{I} \otimes \sigma_z + \sigma_x \otimes \mathcal{I}  - \sigma_x \otimes \sigma_z\Big)
\end{equation*}
\begin{equation*}
U_{\text{gate}}(s=i) = \frac{1}{2}\Big(e^{-\frac{i \pi}{4}}\mathcal{I} \otimes \mathcal{I} + e^{\frac{i \pi}{4}}\mathcal{I} \otimes \sigma_z + e^{\frac{i \pi}{4}} \sigma_x \otimes \mathcal{I}  - e^{-\frac{i \pi}{4}} \sigma_x \otimes \sigma_z\Big)
\end{equation*}
Using Eq.(\ref{E11}) combination gate becomes
\begin{equation}
U_{\text{comp}} = U_{\text{gate}} U_{XOR} = \frac{1}{\sqrt{1+\vert s \vert^2}} \begin{pmatrix}
1 & 0 & 0 & s\\
0 & s^{*} & 1 & 0 \\
s & 0 & 0 & 1 \\
0 & 1 & s^{*} & 0
\end{pmatrix}
\label{E12}
\end{equation}
One applies $U_{\text{comp}}$ gate to the initial state 
\begin{equation}
U_{\text{comp}} \vert \Psi \rangle_{i} = \frac{1}{1+\vert z \vert^2} \frac{1}{\sqrt{1+\vert s \vert^2}} \begin{pmatrix}
z^2 + s \\ z(s^{*} + 1) \\ s z^2 + 1 \\ z(s^{*}+1)
\label{E13}
\end{pmatrix}
\end{equation}
After this one applies post-selection operator ($P$),
\begin{equation}
P = \mathbf{I} \otimes \vert \uparrow \rangle_z \langle \uparrow \vert = \begin{pmatrix}
1 & 0 \\
0 & 1
\end{pmatrix} \otimes \begin{pmatrix}
1 & 0 \\
0 & 0
\end{pmatrix}  = \begin{pmatrix}
1 &  0 & 0 & 0\\
0 & 0 & 0 & 0\\
0 & 0 & 1 & 0\\
0 & 0 & 0 & 0
\end{pmatrix}
\label{E14}
\end{equation}
Applying this post-selection operator on Eq.(\ref{E13}) one gets,
\begin{equation}
P U_{\text{comp}}\vert \Psi \rangle_{i} =P \vert \Psi(1)\rangle  =\frac{1}{1+\vert z \vert^2} \frac{1}{\sqrt{1+\vert s \vert^2}} \begin{pmatrix}
z^2 + s \\
s z^2 + 1
\end{pmatrix} \otimes \vert \uparrow \rangle_z
\label{E15}
\end{equation}
Ultimately by appropriate normalization (postselection) on the first qubit one gets the evolved first qubit state as,
\begin{equation}
\vert \tilde{\psi} \rangle = \dfrac{P \vert \Psi(1)\rangle}{\sqrt{\langle \Psi(1) \vert  P \vert \Psi(1)\rangle}}= \frac{1}{\sqrt{1+\vert f(z) \vert^2}} \begin{pmatrix}
f(z) \\ 1
\end{pmatrix}, 
\label{E16}
\end{equation}
where  $f(z) = \frac{z^2 + s}{s z^2 + 1}$. This is the implementation of QCUs for generating first iterated state $\vert \tilde{\psi} \rangle$. 
\par In the same way taking identical two qubit state $\vert \tilde{\psi} \rangle \otimes \vert \tilde{\psi} \rangle$ as initial state and following the steps from Eq.(\ref{E10}) to Eq.(\ref{E16}) one arrives at the second iterated state $\vert \tilde{\psi} \rangle_1$.
\begin{equation}
\vert \tilde{\psi} \rangle_1 = \frac{1}{\sqrt{1+\vert f^{(2)}(z) \vert^2}} \begin{pmatrix}
f^{(2)}(z) \\ 1
\end{pmatrix}, 
\label{E17}
\end{equation}
By repetition of the above process one eventually arrives at the $n$-th iterated desired state,
\begin{equation}
\vert \tilde{\psi} \rangle_{n-1} = \frac{1}{\sqrt{1+\vert f^{(n)}(z) \vert^2}} \begin{pmatrix}
f^{(n)}(z) \\ 1
\end{pmatrix}, 
\label{E18}
\end{equation}
It is worth pointing out that with successful generation (associated with success probability $p_{\text{success}}$ of each successive desired iterated qubit state, the ensemble of identically prepared qubits reduces by a factor of $2$ as iteration($n$) increases. This is a manifestation of the fact that one really needs to start with $\Big(\frac{2}{p_{\text{success}}}\Big)^n$ no. of identically prepared qubits at the beginning to successfully generate the desired $n$-th iterated qubit state.


\begin{figure}[htb!] 
    \centering
\begin{quantikz}
\lstick{$\ket{\psi}$} & \ctrl{1}\gategroup[2,steps=3,style={dashed,rounded
corners,fill=blue!20, inner
xsep=2pt},background,label style={label
position=below,anchor=north,yshift=-0.2cm}]{{\sc
$U_{\text{comp}}$ gate}} &\gate[2]{U_{\text{gate}}}&  & &
  \rstick{$\ket{\tilde{\psi}}$}\\
\lstick{$\ket{\psi}$} & \targ{} & & & \meter{} &  \rstick{$\ket{{\uparrow}}_z$}
\end{quantikz}
\caption{\textbf{Quantum circuit unit (QCU) of FLNC map:} of the form $f(z) = \dfrac{z^2 + s}{s z^2 + 1}$. In the left of the quantum circuit, a separable state $\ket{\psi} \otimes \ket{\psi}$ is fed as input which generates an output state $\ket{\tilde{\psi}} \otimes \ket{\uparrow}_z$ after post-selection of which the renormalized state of the 1st qubit mimics the evolution generated by FNLC map (Eq.{\ref{E02}}). }.
\label{fig5}
\end{figure}

\sectA{Success Probability of the emulated dynamics induced by FNLC map for $s=i$ \label{SM6}}

In this subsection of supplementary material, we provide the analytical form of success probability distribution for 1st iteration yielding the desired outcome of the emulated FNLC map (Eq.(\ref{E02})) with $s=i$ and also present the average success probability of the first iteration to be successful (yielding the desired evolved state).

\par Following Eq.(\ref{E13}-\ref{E15}) we write the success probability (yielding the state in Eq.(\ref{E15}) in our favor as per our protocol) as,

\begin{equation}
p_{\text{success}} = \langle \Psi(1) \vert \hat{P}^{\dagger}\hat{P} \vert \Psi(1) \rangle = \frac{1+\vert z \vert^4}{(1+\vert z \vert^2)^2}
\label{E23}
\end{equation}

In fact we can replicate the same process for generating the desired $n$-th iterated state with the following success probability,

\begin{equation}
p_{\text{success}}(n) = \langle \Psi(1) \vert \hat{P}^{\dagger}\hat{P} \vert \Psi(1) \rangle = \frac{1+\vert f^{(n-1)}(z) \vert^4}{(1+\vert f^{(n-1)}(z) \vert^2)^2}
\label{E24}
\end{equation}

Given one can always write the initial state correspondent complex number $z$ as $z = \cot{\frac{\theta}{2}}e^{i \phi}$ \footnote{For the FNLC map with $s=i$ (Eq. \ref{E02}), taking $\theta_0 = \theta, \phi_0 = \phi$ one can write $$ \theta_n = 2 \cot^{-1}\Big[\sqrt{\frac{3 + \cos 2 \theta_{n-1} + 2 \sin^2\theta_{n-1} \sin 2 \phi_{n-1}}{3 + \cos 2 \theta_{n-1} - 2 \sin^2\theta_{n-1} \sin 2 \phi_{n-1}}}\Big]$$ and $$ \phi_n = -2 \tan^{-1}\Big[\frac{\cos \theta_{n-1}}{\cos 2 \phi_{n-1} \sin^2 \theta_{n-1}} \Big] $$ where $z_n = f^{(n)}(z) = \cot[\frac{\theta_n}{2}]e^{i \phi_n}$ with $z_n$ being the complex number corresponding to $n$-th iterated state.}, the expression in Eq.(\ref{E24}) simplifies to,

\begin{equation}
p_{\text{success}}(n)= \frac{1}{4} ( 3 + \cos{2 \theta_{n-1}} ) 
\label{E25}
\end{equation}

which in turn gives the \textit{average success probability} as,
\begin{equation}
\langle p_{\text{success}} \rangle =\frac{\int_{\theta} \int_{\phi} p_{\text{success}} \  \sin \theta d \theta d \phi}{\int_{\theta} \int_{\phi} \sin \theta d \theta d \phi } = \frac{2}{3}
\label{E26} 
\end{equation}




\sectA{Roots of FNLC map for $s = i$ :\ $f^{(2n+1)}(z) = 1$ \label{SM7}}

We must point out that the statistical correlation of ensembles of $K_3$ data series is not very generic as shown in FIG.\ref{fig1}\textcolor{blue}{-(b)}. In fact the nature of this fantastic outcome of $r_{XY}$ stabilizing towards zero is the manifestation of the fact that one of the eigenstates of the dichotomic operator is the fixed point of the map situated at $\ket{+x}$ (+ve eigenstate of $\sigma_x$)  upon repeated discrete time evolution. Moreover, the randomness of correlation (as depicted in FIG.\ref{fig1}\textcolor{blue}{-(b)}) is solely because of the chaotic evolution of the initial states chosen from the Julia set.
\par  Upon careful analysis we can say that experimentally it is only enough to measure the expectation value of the particular dichotomic measurement operator with the evolved initial states where eigenstates of the measurement operator satisfy $f^{(2n+1)}(z) = 1$. It turns out that there are several other measurement operator solutions ($\vec{\sigma} \cdot \hat{n}$) which are roots of this equation namely,
\begin{equation*}
\hat{n} = \ -\hat{i}, \ \hat{j}, \ -\hat{j}, \ \frac{1}{\sqrt{2}}(\hat{i} + \hat{j}) , \  \frac{1}{\sqrt{2}} (\hat{i}-\hat{j}), \ \frac{1}{\sqrt{2}}(-\hat{i}+\hat{j}), \ \frac{1}{\sqrt{2}}(-\hat{i}-\hat{j})
\end{equation*}
All these measurement operator will lead to the same qualitative outcome as FIG.\ref{fig1}\textcolor{blue}{-(b)}. It is also a good point to mention that in experiment we do not need to measure the evolution of the measurement operator eigenstates through QCUs once it reaches the fixed point situated at $\ket{+ x}$. This reduces a significant number of QCUs used in the experiment. \\

\end{document}